\documentclass{elsart}

\usepackage{algorithm}
\usepackage{algorithmic}
\usepackage{amsmath}
\usepackage{subfigure}

\newtheorem{definition}{Definition}

\newcommand{\ab}{\allowbreak}
\newcommand{\iold}{\vb{m}}
\newcommand{\cold}{\vb{d}}
\newcommand{\vb}[1]{#1}

\begin{document}

\begin{frontmatter}

\title{Privacy Preserving ID3 over Horizontally, Vertically and Grid Partitioned Data}

\author{Bart Kuijpers}\ead{bart.kuijpers@uhasselt.be},
\author{Vanessa Lemmens},
\author{Bart Moelans}\ead{bart.moelans@uhasselt.be}
\address{Theoretical Computer Science,\\ Hasselt University \& Transnational University Limburg,\\ Belgium}

\author{Karl Tuyls}
\address{Department.of Industrial Design,\\ Eindhoven University of Technology,\\  The Netherlands}
\ead{k.p.tuyls@tue.nl}

\begin{abstract}
We consider privacy preserving decision tree induction via ID3 in the case where the training data is horizontally or vertically distributed. Furthermore, we consider the same problem in the case where the data is \emph{both} horizontally and vertically distributed, a situation we refer to as \emph{grid partitioned data}.
We give an algorithm for privacy preserving ID3 over horizontally partitioned data involving more than two parties. 
For grid partitioned data, we discuss two different evaluation methods for preserving privacy ID3, namely, first merging horizontally and developing vertically or 
first merging vertically and next developing horizontally. Next to introducing privacy preserving data mining over grid-partitioned data, the main contribution of this paper is that we show,  by means of a complexity analysis that the former evaluation method is the more efficient.

\end{abstract}

\end{frontmatter}

\section{Introduction}

\subsection{Privacy preserving data mining}
In recent years privacy preserving data mining has emerged as a very active research area in data mining. The application possibilities of data mining, combined with the Internet, have attracted and inspired many scientists from different 
research areas such as computer science, bioinformatics and economics, to actively participate in this relatively young field. Over the last few years this has naturally lead to a growing interest in security or privacy issues in data mining. More precisely, it became clear that discovering knowledge through a combination of different databases, raises important security issues. Despite the fact that a centralized warehouse approach allows to discover knowledge, which would have not emerged when the sites were mined individually, privacy of data cannot be guaranteed in the context of datawarehousing. 

Although data mining results usually do not violate privacy of individuals, it cannot be assured that an unauthorized person will not access the centralized warehouse with some malevolent intentions to misuse gathered information for his own purposes during the data mining process. Neither can it be guaranteed that, when data is partitioned over different sites 
and data is not encrypted, it is impossible to derive new knowledge about the other sites. Data mining techniques try to identify regularities in data, which are unknown and hard to discover by individuals. Regularities or patterns are to be understood as revelations over the entire data, rather than on individuals. However to find such revelations the mining process has to access and use individual information.

More formally, this problem is recognized as the \textit{inference problem} \cite{Clift96,Csil01}. Originally this problem dates back to research in database theory during the 70s and early 80s, acknowledged back then as access control. Models were developed offering protection against unauthorized use or access of the database. However, such models seemed unable to sufficiently protect sensitive information. More precisely, indirect accesses (through a different database and metadata) still allowed  one to attain information not authorized for. 
Here, metadata \vb{consists}, e.g., of dependencies between different databases, integrity constraints, domain knowledge, etc. In other words, the inference problem occurs when one can obtain vital information through metadata violating individuals (or companies) privacy. With the elaboration of different network technologies and growing interest in pattern recognition this problem naturally carries over to data mining. In fact it gets even worse as illustrated by Sweeny in \cite{Swee01}. In her work Sweeny \vb{shows} that the typical de-identification technique applied on data sets for public use, does not render the result anonymous. More precisely, she demonstrated that combinations of characteristics (or attributes) can  construct a unique or near-unique identifier of tuples, which means that information can be gained on individuals even when their identifiers are distorted.

Over the past few years state of the art research in privacy preserving data mining has concentrated itself along two major lines: data which is \emph{horizontally distributed} and data which is \emph{vertically distributed}. Horizontally partitioned data is data which is homogeneously distributed, meaning that all data tuples yield over the same item or feature set. Essentially this boils down to different data sites collecting the same kind of information over different individuals. Consider for instance a supermarket chain which gathers information on the buying behavior of its customers. Typically, such a company has different branches, implying data to be horizontally distributed. Vertically distributed data is data which is heterogeneously distributed. Basically this means that data is collected by different sites or parties on the same individuals but with differing item or feature sets. Consider for instance financial institutions as banks and credit card companies, they both collect data on customers having a credit card but with differing item sets.

\subsection{Our contribution}
In this paper, we also consider data which is \emph{both} horizontally and vertically distributed, which we will call \emph{grid partitioned data}. To our knowledge, there has been no research up till now in privacy preserving data mining that considers grid distributed data. However this kind of situation seems highly relevant and significant. Consider for instance the situation were different financial institutions gather data on clients concerning savings account, investments, credit cards and others. This situation clearly considers data which is grid partitioned, since some institutions deal with credit cards and not with investments and vice versa and since financial institutions  
typically have data emerging from different branches of a bank.
For a more thorough elaboration of this example, we refer to the end of this Introduction (Section \ref{subsec:motexample}). In this paper, we propose a new algorithm to preserve privacy when constructing  a decision tree for classification over grid partitioned data using ID3, involving multiple parties. Most closely related to this work is that of Lindell and Pinkas \cite{Lind01} who introduced a secure multi-party computation technique for classification using the ID3 algorithm over horizontally partitioned data and that of Du and Zhan \cite{Du02} who introduced a protocol for making ID3 secure over vertically partitioned data. An important contribution of our work is to consider horizontally and vertically distributed data at the same time. Furthermore, we also believe this is highly significant as most real life vital data mining situations, involving multiple parties, consist of grid partitioned data. A motivating example is discussed in Section \ref{subsec:motexample}. For grid partitioned data, we discuss two different evaluation methods for preserving privacy ID3, namely, first merging horizontally and developing vertically or 
first merging vertically and next developing horizontally. We show in Section \ref{sec:discussion} by means of a complexity analysis that the former is the most efficient.  
In the context of 
secure multiparty computation,  the ``semi-honest'' and the ``malicious'' model~\cite{Yao,goldreich} are considered. In the former, 
 all parties follow the protocol strictly, but 
 are allowed to remember everything they encounter while executing the 
protocol and to use this information to compute information about the other parties, whereas in the malicious model the parties are allowed to cheat. We assume the semi-honest model in this paper.

The rest of this paper is structured as follows. We end this Introduction by a motivating example for grid partitioned data, illustrating the importance of efficiently dealing with grid partitioned data in data mining applications. 
In Section \ref{sec:preliminaries}, we sketch the preliminaries as the ID3 algorithm and definitions of horizontally, vertically and grid distributed data. Section \ref{sec:algorithm} introduces our algorithm. We discuss the two different evaluation 
methods for preserving privacy ID3 in Section \ref{sec:discussion}. Section \ref{sec:conclusion} concludes the paper.

\subsection{Motivating example for grid partitioned data}
\label{subsec:motexample}

Typically for financial institutions as banks is that they offer their clients different services as a savings account, choice of credit card, Maestro and all kinds of investment possibilities as mortgages, stock investments, fund orders and so on. Of course a bank is interested in knowing which are good customers, which are bad ones and which are possible defrauders. Reasons are obvious, making profit and avoiding losses because of clients which are not credit worthy and show unreliable behavior. Gathering all kinds of financial data about their customers and their transactions can help them in identifying risky clients and possible defrauders, preventing huge financial losses. More precisely, by using good data mining techniques it becomes possible to generalize over these gathered data sets and identify possible risks for future cases or transactions. Typical for different branches is to gather the same kind (i.e. item sets) of data on different clients, implying that data is \textit{horizontally} partitioned. A possible item set \vb{re-occurring} at banks is illustrated in Table \ref{tab:itemset}.

\begin{table}[ht]
\centering
\caption{A possible item set.}
\begin{tabular}{|c|c|c|c|c|c|c|}
\hline Cust.~nr. & mortgage & account & salary & stock & neg. saldo & Fraudulent? \\ \hline 
$A11$& $25.000$ & $104.200$ & $2.200$ & $0$ &$no$ & no\\
$B12$& $0$ & $1.001.020$ & $4.4000$ & $1.000$ &$yes$ & yes\\
$\vdots$&\vb{$\vdots$}&\vb{$\vdots$}&\vb{$\vdots$}&$\vdots$&\vb{$\vdots$}&$\vdots$\\
\hline
\end{tabular}\label{tab:itemset}
\end{table}

However by combining their data sets, it would become possible to derive knowledge, leading to  a high level of precision in triggering fraudulent behavior, which they would not have reached for individually. Consider for instance a simplified rule $X, Y \rightarrow F$, meaning that if features $X$ and $Y$ are satisfied this implies a high chance the transaction is fraudulent. It is not imaginary that this association rule is known to bank $A$ and unknown to bank $B$, simply because $B$ has not enough (local) tuples to support this rule. There is a reasonable chance that by combining their databases, sites $A$ and $B$ would have discovered association rules which globally hold and which they would have not discovered individually, implying a greater accuracy in identifying defrauders. Although such a cooperative behavior could save them a great deal of money, none of them, as they are competitors, would be willing to share all its transactions and itemsets with one another for obvious reasons.

Although it is possible for banks to gather substantial data on their clients there is still room for more improvement. More precisely, a bank does not typically manage all the services it offers. For instance credit card transactions are managed by separate companies collaborating with banks. Despite this cooperation, neither of them will be too happy to exchange data and item or feature sets on their customers. Still if they would be willing to collaborate, this could lead to a higher precision in identifying fraudulent cases and all parties would benefit. In other words, the group of people having a credit card is usually involved in an investment of all possible kinds: mortgages, stock market, order funds etc. This implies that the group of individuals on which the credit card companies gather data is more or less the same as the group on which banks gather data concerning investments. This boils down to data which is \textit{vertically} partitioned. 

Summarizing, this example shows that it is not imaginary at all that data appears to be as well horizontally as vertically distributed in real life situations, which we call \textit{grid partitioned data}. Note that we can easily extend the above example to contain more parties. We could for instance add tax services, interested in tracking people cheating on their taxes. Most importantly, the example illustrates that we do not only need to consider privacy preserving techniques for horizontally or vertically distributed data, but that it is highly significant for real life applications to consider the combination of both.

\section{Preliminaries}\label{sec:preliminaries}
We start this section with a subsection summarizing the ID3 algorithm. Then we continue with a  subsection describing definitions and examples of horizontally, vertically and grid distributed data. We continue with preliminaries on multi-party computation.
 
\subsection{The ID3 algorithm}\label{subsec:ID3}
The ID3 algorithm (Inducing Decision Trees) was originally introduced by Quinlan in \cite{Quin86} and is described below in Algorithm \ref{algo:id3}. Here we briefly recall the steps involved in the algorithm. For a thorough discussion of the algorithm we refer the interested reader to \cite{Mit97}. 

The input of ID3 is a finite data set of tuples containing (discrete or nominal) values for a finite number of attributes, one of which is called the class attribute (also called target class). ID3 induces a decision tree from an example set in a top-down manner. More precisely, the algorithm starts at the root node, choosing each time the attribute which separates the data most efficiently according to their target class. Then the algorithm creates a branch for each value of this attribute and continues from there by repeating the above process until all attributes are used. To determine which attribute is best in classifying the given data set, a measure from information theory is used, namely \textit{information gain}.  Information gain is defined as the expected reduction in \textit{entropy}. Entropy measures the homogeneity of a data set.  More formally, the entropy of a data set of tuples $S$ is defined as:
\begin{equation}
entropy(S) = \sum_{i=1}^{\cold} -p_i log_2 p_i
\end{equation}

\noindent where $\cold$ is the total number of different values the target class can take on and $p_i$ is the proportion of tuples of the data set  having target value $i$. The information gain of an attribute $A$ is then defined as:
\begin{equation}
gain(S,A)=entropy(S) - \sum_{v} \frac{|S_v|}{|S|} entropy(S_v)
\end{equation}

\noindent with $S_v$ the subset of S with tuples having value $v$ for attribute $A$.

\begin{algorithm}[ht]
\caption{The ID3 Algorithm}
\label{algo:id3}
\begin{algorithmic}[1]
\REQUIRE $R$, a set of attributes.
\REQUIRE $C$, the class attribute. 
\REQUIRE $S$, data set of tuples.
\IF{$R$ is empty}
\STATE Return the leaf having the most frequent value in data set $S$. 
\ELSIF{all tuples in $S$ have the same class value} 
\STATE Return a leaf with that specific class value. 
\ELSE
\STATE Determine attribute $A$ with the highest information gain in $S$. 
\STATE Partition $S$ in $m$ parts $S(a_{1}), ..., S(a_{m})$ such that $a_{1}, ..., a_{m}$ are the different values of $A$. 
\STATE Return a tree with root $A$ and $m$ branches labelled $a_{1} ... a_{m}$, such that branch $i$ contains ${\rm ID3}(R - \left\{A\right\},C,S(a_{i}))$. 
\ENDIF
\end{algorithmic}
\end{algorithm}

\subsection{Horizontally, vertically and grid partitioned data}
In this section we provide a formal definition of horizontally, vertically and grid partitioned data. We will use the projection operation as defined in relational algebra in database theory.

Suppose we have:
\begin{enumerate}
\item A  relation (or data set) $S$ over the schema $I, A_1,...,A_{\vb{|R|}},C$ consisting of a finite number of tuples. The attribute $I$ is supposed to be a key (i.e., contain identifiers) and is not considered as an attribute to build the decision tree. The only purpose of 
the attribute $I$ is to be able to join vertically distributed data. The attribute $C$ will be \vb{referred} to as the class attribute.
\item Parties $P_{ij}$ with $i=1,..,\vb{v}$,  $j=1,..,\vb{h}$ and $\vb{v}$ smaller than the number of attributes (i.e., $\vb{|R|}+1$) 
\item Each party $P_{ij}$ is holding a part $S_{ij}$  containing information about certain attributes (including $I$) and certain tuples. The $S_{ij}$ are such that 
\begin{itemize}
\item $S_{ij}$ is a partition of $S$, more precisely $S=\cup_{j=1}^{\vb{h}} \bowtie_{I, i=1}^{\vb{v}} S_{ij}$;
\item $S_{ij}$ and $S_{ij'}$ have the same attributes but (parts of) different tuples of $S$ when $j\not= j'$;
\item $S_{ij}$ and $S_{i'j}$
have disjoint attributes but contain information about the same tuples of $S$.
\end{itemize}
\end{enumerate}

\begin{definition} We call  $S$  
\begin{itemize}
\item \emph{horizontally distributed} if and only if $\vb{v}=1$;
\item \emph{vertically distributed} if and only if $\vb{h}=1$; and
\item \emph{grid distributed} if and only if $\vb{v,h}\geq 2$.
\end{itemize}
\end{definition}

\begin{figure}[ht]
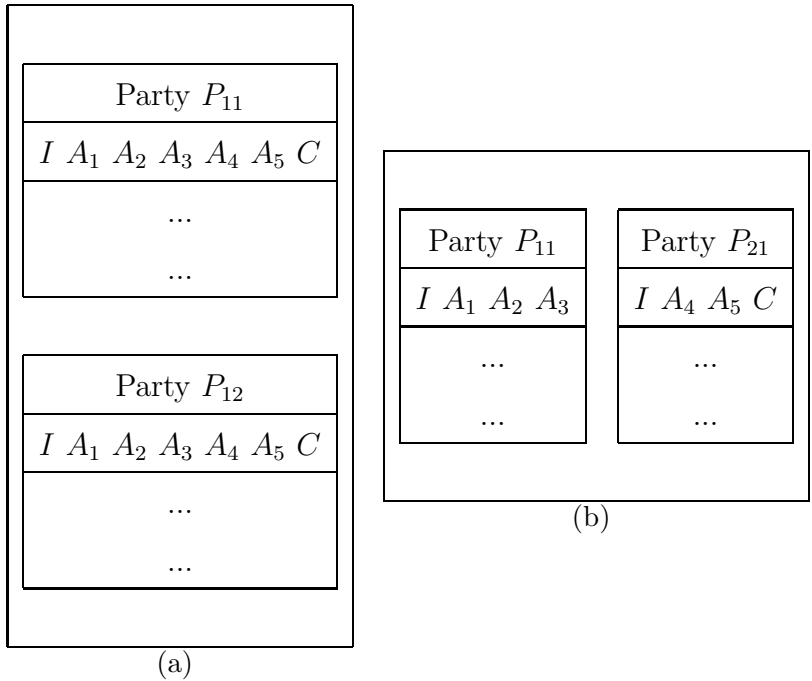

 \centering{
\subfigure[]{
\begin{tabular}{|c|}
\hline
\phantom{joske}\\
\begin{tabular}{|c|}
\hline
Party $P_{11}$\\
\hline
$I$\ $A_1$\ $A_2$\ $A_3$\ $A_4$\ $A_5$\ $C$\\
\hline
...\\
...\\ 
\hline
\end{tabular}
\\
\phantom{joske}\\
\begin{tabular}{|c|}
\hline
Party $P_{12}$\\
\hline
$I$\ $A_1$\ $A_2$\ $A_3$\ $A_4$\ $A_5$\ $C$\\
\hline
...\\
...\\ 
\hline
\end{tabular}\\
\phantom{joske}\\
\hline
\end{tabular}
\label{fig:hor-data}}
\subfigure[]{
\begin{tabular}{|c c|}
\hline
\phantom{joske}& \\
\begin{tabular}{|c|}
\hline
Party $P_{11}$\\
\hline
$I$\ $A_1$\ $A_2$\ $A_3$\\
\hline
...\\
...\\ 
\hline
\end{tabular} &
\begin{tabular}{|c|}
\hline
Party $P_{21}$\\
\hline
$I$\ $A_4$\ $A_5$\ $C$\\
\hline
...\\
...\\ 
\hline
\end{tabular}\\
\phantom{joske}&\\
\hline
\end{tabular}
\label{fig:ver-data}}
}
\caption{(left) Horizontally distributed data.(right) Vertically distributed data.}
\end{figure}

\rm
Examples of horizontally, vertically and grid distributed databases can be found in Figures \ref{fig:hor-data}, \ref{fig:ver-data} and \ref{fig:grid-data}.

\begin{figure}[ht]
\centering{
\begin{tabular}{|c c c|}
\hline
\phantom{joske}& &\\
\begin{tabular}{|c|}
\hline
Party $P_{11}$\\
\hline
$I$\ $A_1$\ $A_2$\\
\hline
...\\
...\\ 
\hline
\end{tabular} &
\begin{tabular}{|c|}
\hline
Party $P_{21}$\\
\hline
$I$\ $A_3$\\
\hline
...\\
...\\ 
\hline
\end{tabular} &
\begin{tabular}{|c|}
\hline
Party $P_{31}$\\
\hline
$I$\ $A_4$\ $A_5$\ $C$\\
\hline
...\\
...\\ 
\hline
\end{tabular}\\
\phantom{joske}&&\\
\begin{tabular}{|c|}
\hline
Party $P_{12}$\\
\hline
$I$\ $A_1$\ $A_2$\\
\hline
...\\
...\\ 
\hline
\end{tabular} &
\begin{tabular}{|c|}
\hline
Party $P_{22}$\\
\hline
$I$\ $A_3$\\
\hline
...\\
...\\ 
\hline
\end{tabular} &
\begin{tabular}{|c|}
\hline
Party $P_{32}$\\
\hline
$I$\ $A_4$\ $A_5$\ $C$\\
\hline
...\\
...\\ 
\hline
\end{tabular}\\
\phantom{joske}&&\\

\begin{tabular}{|c|}
\hline
Party $P_{13}$\\
\hline
$I$\ $A_1$\ $A_2$\\
\hline
...\\
...\\ 
\hline
\end{tabular} &
\begin{tabular}{|c|}
\hline
Party $P_{23}$\\
\hline
$I$\ $A_3$\\
\hline
...\\
...\\ 
\hline
\end{tabular} &
\begin{tabular}{|c|}
\hline
Party $P_{33}$\\
\hline
$I$\ $A_4$\ $A_5$\ $C$\\
\hline
...\\
...\\ 
\hline
\end{tabular}\\

\phantom{joske}& &\\
\hline
\end{tabular}
   \caption{Grid distributed data.}
    \label{fig:grid-data}}
\end{figure}

\subsection{Preliminaries on multiparty computation}
\label{subsec:smpc}
In this section we recall some results from multiparty computation that will be needed
as building blocks in the algorithms in the next section.

Basically, secure multi-party computation (SMPC) makes sure that different parties involved in a computation process, do not learn anything more than the result(s) of the computation process and anything else that is derivable  in a polynomial amount of time
(without cheating). More precisely, SMPC is of great interest to  the inference problem. In the case of horizontally, vertically and grid partitioned data in data mining, the mining process requires a lot of communication between the different parties. The SMPC techniques prevent  any party from deriving new knowledge about the other parties involved.

It is not our intention to give a complete overview here of SMPC, therefore we refer to \cite{lindell,Clift01,Yao,Lind01,Chun01,Agra00,Clift96,vaidya}. Here we provide the security protocols necessary for our purposes, i.e., 
\begin{itemize} 
\item the secure sum protocol, 
\item the Yao circuit, 
\item the secure union protocol, 
\item the secure size of set intersection  protocol; and 
\item the $xln(x)$ protocol.
\end{itemize}

We remark that in the context of 
secure multiparty computation, two models, that we already mentioned before, are considered, namely the ``semi-honest'' and the ``malicious'' model~\cite{goldreich,Yao}.

In the \emph{semi-honest model}, all parties follow the protocol strictly.
They are allowed to remember everything they encounter while executing the 
protocol and to use this information to compute (in polynomial time) information about the other parties.
In the \emph{malicious model} the parties are allowed to cheat.
They may for example falsify their inputs in order to learn more about the input of other parties.

We will assume the semi-honest model in the description of our algorithms.

\subsubsection{Secure sum protocol}
The goal of this protocol is that $k>2$ parties can compute the sum of the values each party holds in such a way that no party can learn anything about the values of the other parties. 

The protocol of Kantarcioglu and Clifton~\cite{kant02}
protects individual values by using a random number. Party $0$ adds a random number to its own value and sends it to Party $1$. Party $1$ cannot learn anything from this value due to the random number. Party $1$ adds his value to this number and sends it along to Party $2$. This process continues until the last party has been reached. This party adds his number to the number it received and sends it to Party $0$. Party $0$ can now compute the sum by distracting the random number from the sum it received of the last party. Now Party $0$ reveals the sum to the other parties.

\newpage
How safe is this protocol? It can be shown, by means of a polynomial time simulator, 
that, in the semi-honest model, this protocol is safe. Actually, to show safety it is necessary that all values remain within a finite domain $[0,m]$ and all computations are done modulo $m$.
 
It should be remarked that the protocol can be broken if we assume the malicious model. For instance, it is clear that when Party $i-1$ and Party $i+1$ collaborate, the value of Party $i$ can be discovered. As a remedy for this problem, each party can split its value in $n$ parts. Of all parts the sum is calculated. To avoid that  
   Party $i-1$ and Party $i+1$ can collaborate to discover the value of Party $i$, during each of these $n$ computations different paths are followed.
   In this way, more parties have to collaborate to discover individual values.
    
\subsubsection{Yao circuit}
Yao introduced in~\cite{Yao} the concept of \emph{secure two party computation}.
He showed that any function $f(x, y)$, where $x$ is the input of Party 1 and $y$ the input of Party 2, can be evaluated in a secure way.

To formalize the concept of security, we concentrate on functions $f$ (Yao makes use of Boolean circuits to represent a function $f$) of the form 
$f(x, y) =
(f_1(x, y), f_2(x, y)).$ This function receives a part of its input, namely $x$, 
from Party 1 and the other part of its input, namely $y$, from Party 2.
Party 1 wants to learn $f_1(x, y)$
and Party 2 wants to learn $f_2(x, y)$.
Suppose that protocol $\Pi$ is used to learn $f$.
$View^{\Pi}_i$ is what Party $i$ learns by executing protocol $\Pi$ and 
$Output^{\Pi}_i$ is the output of Party $i$ ($i=1,2$).
Finally, let $S_i$  be an algorithm that can be executed in polynomial time.
 Yao defines 
$$\{ S_1(x,f_1(x,y)), \ab f_2(x,y))\}=\{View_1^\Pi (x,y), \ab Output_2^\Pi (x,y))\}$$
and
$$\{(f_1(x,y),\ab S_2(y,f_2(x,y)))\} =\{Output^\Pi(x,y),\ab View^\Pi_2 (x,y))\}$$
 meaning that any party can learn from $f$, by executing protocol $\Pi$, 
 only those facts that can be learned in polynomial time from his/her input and
 his/her output.
 Executing the protocol does therefore not provide any extra information.
 
 We remark that Goldreich \emph{et al.}~\cite{goldreich} generalized the results of Yao to more than two parties. Goldreich \emph{et al.} also gave the composition theorem
that states that if a  function $g$ can be reduced safely to a function $f$, and if there is a protocol to safely compute $f$, then also $g$ can be computed safely.

In this paper, we will refer to this type of circuits as \emph{Yao circuits}, even if they concern more than two parties.

\subsubsection{Secure union protocol}
When there are only two parties, computing the union of two sets belonging to
each of those parties, this can lead to security problems. Indeed, the knowledge about 
ones own set and about the union, gives (at least partial) knowledge about the other parties set. In this section, we outline a method to compute the union of $k$ itemsets, belonging to $k$ parties, $k>2$. The goal is that all parties should learn the union, without learning about the itemsets of other parties.
The algorithm is from Kantarcioglu and Clifton~\cite{kant02} and consists of four phases that we sketch below. These authors also show its security.

\par\noindent \emph{Phase 1}: All parties generate a commutative, deterministic \vb{encryption} key $E_i$
and a decryption key $D_i$. Each itemset is augmented with fake or dummy items
(this is done to prevent the determination of the cardinality of the itemset).
At the end
of Phase 1, each party has an itemset of the same size (which is agreed upon at the start).

\par\noindent \emph{Phase 2}: Each party encrypts its items and communicates them to the next party
(the communication is cyclic as in the case of secure sum computation, i.e., Party $i$ sends 
information to Party $(i+1)\mod k$). Each party encrypts what he receives and passes it to the next party. This continues until each Party $i$ is in the possession of the completely encrypted items of Party $(i+1)\mod k$. We remark that to 
continue one more step 
would be no longer secure.

\par\noindent \emph{Phase 3}: 
The even-numbered parties send all items in their possession to Party $0$ and the 
odd-numbered parties do the same to Party 1 (the last party always has to send to Party 1 to avoid that a party gets its own fully encrypted itemset).
Parties 0 and 1 take the union of what they received and remove the doubles.
Party 1 sends everything he has to Party 0, who removes the doubles.
At this point the union is in the possession of Party 0, \vb{be it} fully encrypted.

\par\noindent \emph{Phase 4}: The encrypted union is sent to all parties to be decrypted.
Finally the fake items are removed and the result is announced to all
parties.

\subsubsection{Secure size of set intersection protocol}
When there are only two parties, computing the size of the intersection of two sets belonging to
each of the parties can lead to security problems. Indeed, the knowledge about 
ones own set and about the size of the intersection of two sets  gives (at least partial) knowledge about the set of the other party. So, we are interested to 
compute the size of set intersection of $k$ itemsets, belonging to $k$ parties, $k>2$. 
The goal is that all parties should learn the size of set intersection, without learning about the itemsets of other parties.

\newpage
Jaideep Vaidya~\cite{vaidya} proposed a protocol for the secure computation of the 
size of set intersection. It is similar to the secure union protocol, and we will not repeat the details here but refer to~\cite{vaidya}.

\subsubsection{$x\ln(x)$ protocol}
The $x\ln(x)$ protocol, due to Lindell and Pinkas~\cite{lindell}, is different from the previous protocols. It uses Yao circuits, as mentioned earlier in this section. Because circuits are only suitable for two parties, 
also this protocol is only suitable for two parties.
Assume we have two parties, called Alice and Bob. Alice has a value 
$x_a$ and Bob has a value $x_b$.
The goal of the $x\ln(x)$ protocol is to give  Alice and Bob both a share 
$s_a$ and $s_b$ respectively, such that $s_a + s_b = (x_a + x_b)\ln(x_a + x_b)$

The $ x\ln(x)$ protocol makes use of two subprotocols. 
The first receives two values $x_a$ and $x_b$ as input and 
returns two random shares of $
\ln(x_a +x_b)$ as output (using a Taylor series). The second, called the \vb{multiplication} protocol, receives two values  $u_a$ and $u_b$ as
input and returns two random shares of  $u_a.  u_b$ as output. 

Alice and Bob run the $\ln(x)$ protocol and become shares
$u_a$ and $u_b$. Next, the multiplication protocol is executed twice.
First with $u_a$ and $x_b$ as input. This gives Alice and Bob respectively shares $v_a$ and $v_b$.
the second time it is called with $x_a$ and $u_b$, giving Alice and Bob
respectively shares $w_a$ and $w_b$.
Alice now has $x_a$, $u_a$, $v_a$ and $w_a$, with which she can compute $s_a = x_a  u_a + v_a + w_a$.
Bob can construct $s_b = x_b  u_b + v_b + w_b$ in a similar way.
Since  $x_a  u_a + x_b  u_b + x_a  u_b + x_b  u_a 
\ab = (x_a + x_b)(u_a + u_b) =\ab 
(x_a + x_b)\ln(x_a + x_b)$, Alice and Bob both have their share of $(x_a + x_b)\ln(x_a + x_b)$.

\section{Privacy preserving ID3: Grid partitioned data}\label{sec:algorithm}
In the present section, we introduce our algorithms, preserving privacy over grid partitioned data. Basically, we will study the following \vb{dilemma}: when data is grid partitioned we can first merge it horizontally and then further develop the process vertically, or the other way around. Obviously other ways  of doing this are possible as well, but we consider only the two straightforward ones. Of course while building the decision tree we need to preserve privacy and use some well known protocols for this. 

In this paper we consider privacy as protecting individual data tuples as well as protecting attributes and  values of attributes. So each party will reveal as little as possible about its data while still constructing an applicable distributed decision tree. The only thing that is known about the tree by all parties is its structure and which party is responsible for each decision node. More precisely, which party possesses the attribute used to make the decision, but not which attribute (and value). We assume that only a limited number of parties know the class attribute and no party knows the entire set of attributes, which is obvious as we use grid partitioned data.

Once the tree is constructed instance classification proceeds as follows. The party that wishes to classify a new unseen instance knows the root node of the tree: or the node resides at his site or he knows the root node-identification (nodeID). A root node identification contains a code \vb{identifying} the party possessing that particular node. Basically, when classifying a new instance, control passes from party to party, depending on the decision nodes that are visited. Every party knows the tuples attribute values for the nodes at its site but knows nothing about the other attribute values. The classification then happens as in Algorithm \ref{classify}.
\begin{algorithm}
\caption{The classification algorithm called \textit{classify(t,nodeID)}. A site wishes to classify a new instance $t$. Control starts at the root node (which every party knows.)}
\label{classify}
\begin{algorithmic}[1]
\IF{The nodeID is a leaf node}
 \STATE its classification value (or distribution) is returned.
\ELSIF{ The nodeID is an interior node}
\STATE node = local node with nodeID
\STATE value = value of attribute node.A (used as decision attribute) for the tuple $t$ we are classifying
\STATE childID = node.value
\STATE return childID.classify(t,childID)
\ENDIF
\end{algorithmic}
\end{algorithm}

Before introducing our new algorithms for grid partitioned data we introduce a minor side result which has not been dealt with so far in the literature, i.e. horizontally partitioned data with more than two parties.

\subsection{Privacy preserving ID3 over Horizontally partitioned data involving more than two parties}
In this section we extend the result of Lindell and Pinkas \cite{Lind01}, i.e. preserving privacy for decision tree learning with two parties, to more than two parties. 
\newpage
Recall the ID3 algorithm from Section \ref{subsec:ID3}. We will separately consider its three basic steps, i.e. \textit{emptiness test} of the attribute set $S$, all transactions having the same class label, i.e. \textit{class label test} and the \textit{default case}. We explain in detail how the algorithm preserves privacy in each of them.

\paragraph{Emptiness test} Thanks to the horizontal distribution of data and the fact that all parties know the intermediate tree, they can easily determine whether $S$ is empty. In case $S$ is empty they have to determine the most frequent class value. This can be calculated by using the \textit{secure sum protocol} for each class value. Each party inputs the number of tuples having the particular class value at his data site to the protocol. In this way they can safely compute for each class value the total number of tuples over all sites, having this value for its class attribute. Now a leaf node can be constructed containing the most frequent class value.\\

\paragraph{Class label test} To securely determine whether all tuples in $S$ have the same class value, a variant of the secure union protocol can be adopted. More precisely, all parties provide \emph{a value} as input to the protocol. If a party has only one class value in all of its tuples, it provides this value as input. Otherwise, a fixed symbol $\bot$ is provided as input.\\
The protocol then runs analogously to the standard secure union protocol until the step in which data has to be decrypted. The first party, i.e. party $0$, has all the values in its possession of which he deletes all doubles. Now there are two possibilities, only one value remains or not. In case of the former, this must be the value which was provided as input by party $0$. This is a class value or the $\bot$ symbol. If it is a class value this means that all parties have provided the protocol with this same value, otherwise it means that all parties still have more than one class value. In case of the latter it is sure that there are still more than one class value. The protocol then has to be stopped, else values are learned by different parties which they should not learn.

\paragraph{Default case} In this case should be determined which attribute classifies data tuples in $S$ most accurately. To calculate information gain, $xln(x)$ has to be calculated a number of times with $x$ partitioned over the different parties (being data tuples). This can be solved by the secure sum protocol. The different parties provide as input their share of $x$ to the protocol. The sum known to the last party and the random value of the first party multiplied by $-1$ are input of the $xln(x)$ protocol. This protocol will provide shares of the result as output, which then can be used as input to a circuit which securely computes it sum and outputs which attribute classifies the tuples best.\\

The complete description can be found in Algorithm \ref{dectreesnieuw}.

\begin{algorithm}
\caption{The privacy preserving ID3 algorithm for more than two parties over horizontally distributed data}
\label{dectreesnieuw}
\begin{algorithmic}[1]
\REQUIRE $R$, The set of attributes.
\REQUIRE $C$, The class attribute. 
\REQUIRE $S$, the horizontally distributed data set. 
\IF{The parties test if $R$ is empty} 
\STATE Secure sum protocol is used to calculate which class value $c_{i}$ is most frequent.

\STATE Return a leaf with class value $c_{i}$. 
\ELSIF{All parties use secure union protocol to test if all tuples in $S$ have the same class value $c_{i}$.} 
\STATE Return a leaf with class value $c_{i}$.  
\ELSE 
\STATE Determine attribute $A$, classifying most accurately tuples in $S$: use secure sum and $xln(x)$ protocols. 
\STATE Partition $S$ in $m$ parts $S(a_{1}), ..., S(a_{m})$ such that $a_{1} ... a_{m}$ are the different values of $A$.
\STATE Return a tree with root $A$ and $m$ branches $a_{1} ... a_{m}$ such that branch $i$ contains $ID3(R - \left\{A\right\},C,S(a_{i}))$. 
\ENDIF
\end{algorithmic}
\end{algorithm}

\subsection{Grid Partitioned data}
We will introduce the grid partitioned privacy preserving algorithm by running through the different steps of ID3 informally. It is important to realize that no site knows the complete attribute set $S$ and only a limited number of parties know the class attribute, more particularly as much as there are horizontal distributions. Note that these algorithms only consider the cases for which the parties \vb{are} denoted by $P_{ij}$ with $i=1,..,\vb{v}$,  $j=1,..,\vb{h}$.


\subsubsection{Horizontal merge and vertical development}

Recall the ID3 algorithm from Section \ref{subsec:ID3}. We will separately consider its three basic steps, i.e. \textit{emptiness test} of the attribute set $S$, all transactions having the same class label, i.e. \textit{class label test} and the \textit{default case}. Here we consider the case that we first merge the data horizontally and continue vertically. A horizontal merge means that we eliminate the horizontal distribution, leaving only a vertical distribution.

\paragraph{Emptiness test} 
To determine if there are any attributes left, as many parties as there are vertical distributions need to cooperate with one another as we need to know all attributes to compute this test. This can be easily understood from Figure \ref{fig:grid-data}. More precisely, in the example of that figure, parties  $Party_{11}$, $Party_{21}$ and $Party_{31}$ can determine together if there are any attributes left. These parties check how many possible attributes they still possess as a candidate decision node and pass this value as input to the secure sum protocol. At the end of the protocol the sum and the  random value are passed to a Yao circuit, which tests if the sum equals zero (meaning that $S$ is empty) or not.\\
In case of the sum being zero, the most frequent class value has to be determined. This is done in the following manner: first all parties determine the tuples reaching the current node in the tree. Then these  tuples are merged horizontally by constructing a union over the vertical groups (over index $i$), i.e. for each vertical group a secure union protocol is applied. In case of Figure \ref{fig:grid-data} we have the following groups computing a union:   $Party_{11}$, $Party_{12}$ and $Party_{13}$; $Party_{21}$, $Party_{22}$ and $Party_{23}$ and $Party_{31}$, $Party_{32}$ and $Party_{33}$.

 Parties which are located on the same horizontal layer (meaning that they have the same index for $j$), which are vertically distributed, will use the same encryption key to compute the vertical unions (i.e. the horizontal merge). In our example these are, $Party_{11}$, $Party_{21}$ and $Party_{31}$; $Party_{12}$, $Party_{22}$ and $Party_{32}$;  $Party_{13}$, $Party_{23}$ and $Party_{33}$. At this stage there is only a vertical distribution left over the entire distributed database, as we \textit{merged} data horizontally. Now we will continue by \textit{developing vertically}. 
The intersection of these different sets with the tuples in a particular class give the number of tuples that reach that point in the tree. This can be done for each class value; note that it is not necessary to use the secure size of set protocol because the unions are already encrypted. This gives us eventually the most frequent class value. Note that it is not necessary to decrypt the values again to compute the intersections. The reason is that we used the same encryption keys for parties at the same horizontal level, implying that equal values in the encrypted unions are also equal in the real unions. Now a leaf can be constructed with a certain leaf identifier. The value of the leaf is known by the parties that have the class attribute. The others only know the identifier.

\paragraph{Class label test} 
Checking whether all transactions in the training set $S$ have the same class value happens analogously to determining the most frequent class value. More precisely, one party knowing the class attribute, can compute the possible intersections, i.e.,  the intersections of the sets of tuples which might reach the current level of the tree or node of the tree with the tuples in a particular class. If all intersections equal zero besides one, all tuples in $S$ have that particular class value. Since they are all the same, now a leaf node is constructed. The parties which were joined in one vertical group knowing the class attribute, all know the id of the node ($nodeID$ in the algorithms) and the specific class value. All other parties just get to know the nodeID of the node, which they need in case they need to classify a new tuple leading to this node in the tree.

\paragraph{Default case} 
In this case, the best classifying attribute has to be determined. To do this, transactions or tuples need to be counted. To learn these numbers recall that entropy and information gain where defined as follows: $entropy(S) = \sum_{i=1}^{\cold} -p_i log_2 p_i$ 
\noindent where $\cold$ is the total number of different values the target class can take on and $p_i$ is the proportion of tuples of the data set  having target value $i$, i.e. $\frac{N_i}{N}$, where $N$ is the total number of tuples reaching the current node and $N_i$ is the number of tuples with attribute value $a_i$. The information gain of an attribute $A$ is then defined as: $gain(S,A)=entropy(S) - \sum_{v} \frac{|S_v|}{|S|} entropy(S_v)$

For each attribute information gain needs to be computed. First all parties determine the tuples reaching the current node in the tree, i.e. $N$. Then these  tuples are merged horizontally by constructing a union over the vertical groups, i.e. for each vertical group a secure union protocol is applied. In case of Figure \ref{fig:grid-data}, we have the following groups computing a union:   $Party_{11}$, $Party_{12}$ and $Party_{13}$; $Party_{21}$, $Party_{22}$ and $Party_{23}$; $Party_{31}$, $Party_{32}$ and $Party_{33}$. For every vertical group, one party will iterate over its attributes. For each such attribute, numbers of tuples need to be counted for every value of this attribute. Thus for every value of the attribute (which is encrypted), an intersection is computed over all the sets, resulting from the horizontal merge. When we know all these numbers, the information gain of this attribute can be computed. This step is repeated for all attributes. The process described so far is called \textit{vertical development}. On of the parties of each vertical group saves the the information gain of its attributes. These parties can then cooperate to compute the best classifying one.  Finally, the party which possesses the best classifying attribute constructs a decision node with is given a node identifier \textit{nodeID}. The value of the node is communicated to the other parties that also possess this attribute. The other parties only get to know the identifier.

The complete description can be found in Algorithm \ref{dectreeshormerge}.

\begin{algorithm}
\caption{The privacy preserving ID3 algorithm over grid partitioned data when data is merged horizontally and further developed vertically.}
\label{dectreeshormerge}
\begin{algorithmic}[1]
\REQUIRE $R$, The set of attributes distributed among the parties $P_{ij}$ with $i=1,..,\vb{v}$,  $j=1,..,\vb{h}$. 
\REQUIRE $C$, The class attribute with $d$ class values, $c_1,...,c_d$. 
\REQUIRE $S$, the grid distributed data set over parties $P_{ij}$ with with $i=1,..,\vb{v}$,  $j=1,..,\vb{h}$ and parties $P_{\vb{v},j}$ holding the class attribute . 
\IF{(Emptiness test)The parties test if $R$ is empty} 
\STATE Secure sum protocol and Yao circuit are used to test whether $R$ is empty.
\STATE In case the attribute set is empty, Secure union protocol is used to merge data horizontally. For the vertical development, the secure size of set intersection protocol does NOT have to be used. A Yao circuit is used to calculate which class value $c_{i}$ is most frequent. A leaf node with class value $c_{i}$ is returned.

\ELSIF{All parties test whether all tuples have the same class value $c_{i}$} 
\STATE Secure union protocoland Yao circuit are used to calculate this.
\STATE In case the test is TRUE, a leaf with class value $c_{i}$ is returned.  
\ELSE 
\STATE Determine attribute $A$, classifying most accurately tuples in $S$: use secure union and secure sum protocols. 
\STATE Partition $S$ in $m$ parts $S(a_{1}), ..., S(a_{m})$ such that $a_{1} ... a_{m}$ are the different values of $A$.
\STATE Return a tree with root $A$ and $m$ branches $a_{1} ... a_{m}$ such that branch $i$ contains $ID3(R - \left\{A\right\},C,S(a_{i}))$. 
\ENDIF
\end{algorithmic}
\end{algorithm}

\subsubsection{Vertical merge and horizontal development}

\paragraph{Emptiness test} 
To determine if there are any attributes left, as many parties as there are vertical distributions need to cooperate with one another as we need to know all attributes to compute this test. So essentially this is done in the same manner as with the horizontal merge, i.e. the previous algorithm.

\newpage
To determine the most frequent class value we will merge data vertically. More precisely, every party first determines the number of tuples that reach the current level of the tree or node of the tree. For this the parties only use the attributes they possess. Then we merge vertically by letting cooperate the parties at the same horizontal level. In our example these are $Party_{11}$, $Party_{21}$ and $Party_{31}$; $Party_{12}$, $Party_{22}$ and $Party_{32}$; $Party_{13}$, $Party_{23}$ and $Party_{33}$. The parties that possess the class attribute now need to compute a set per class value. In our example these are parties $Party_{31}$, $Party_{32}$ and $Party_{33}$. They compute as many secure size of set protocols as there are class values. In this manner they compute per horizontal group (or vertical merge) the number of transactions per class value. If the parties possessing the class attributes have computed these intersections, they have to cooperate to find out the total number of tuples per class value. They compute this by using a secure sum protocol per class value. Then these values are passed on to a Yao circuit to be able to learn the most frequent class value. Now a leaf can be constructed with a certain leaf identifier. The value of the leaf is known by the parties that have the class attribute. The others only know the identifier.

\paragraph{Class label test} 
Determining whether all tuples have the same class value is analogous to the previous step. The difference lies in the Yao circuit, which will test if all sums equal zero except for one. Again a leaf can be constructed with a certain  leaf identifier. The value of the leaf is known by the parties that have the class attribute. The others only know the identifier.

\paragraph{Default case} 
We need to compute the best classifying attribute. To do this, transactions or tuples need to be counted. To learn these numbers recall that entropy and information gain where defined as follows: $entropy(S) = \sum_{i=1}^{\cold} -p_i log_2 p_i$ 
\noindent where $\cold$ is the total number of different values the target class can take on and $p_i$ is the proportion of tuples of the data set  having target value $i$, i.e. $\frac{N_i}{N}$, where $N$ is the total number of tuples reaching the current node and $N_i$ is the number of tuples with attribute value $a_i$. The information gain of an attribute $A$ is then defined as: $gain(S,A)=entropy(S) - \sum_{v} \frac{|S_v|}{|S|} entropy(S_v)$

First, we merge data vertically. More precisely, every party first determines the number of tuples that reach the current level or node of the tree. For this the parties only use the attributes they possess. Then  data is merged vertically by parties at the same horizontal layer (having the same $j$ index in $P_{ij}$) via a secure size of set intersection protocol to obtain exactly those tuples that are in the current dataset associated to the node under consideration in the tree. In our example these are $Party_{11}$, $Party_{21}$ and $Party_{31}$; $Party_{12}$, $Party_{22}$ and $Party_{32}$; $Party_{13}$, $Party_{23}$ and $Party_{33}$. Through a secure sum protocol these numbers can now be added to learn the number of tuples that reach the current node of the tree, which is denoted by $N$ in the entropy formula.

Now we need to compute for each remaining attribute its information gain. This is done by computing for each value of an attribute over each horizontal layer, the number of tuples having this attribute value (we compute $N_i$). This is done by using the secure size of set protocol. Then these numbers can be added over all horizontal layers by using a secure sum protocol. This is called \textit{horizontal development}.  The secure sum (added with the random value) and the random value itself multiplied by one are provided to the $xln(x)$ protocol. This circuit will then output shares of the result which then can be used as input to a circuit which securely computes it sum and outputs which attribute classifies the tuples best. A decision node can now be constructed for the best classifying attribute.

The description of the algorithm is summarized in Algorithm \ref{dectreesvermerge}.

\begin{algorithm}
\caption{The privacy preserving ID3 algorithm over grid partitioned data when data is merged vertically and further developed horizontally.}
\label{dectreesvermerge}
\begin{algorithmic}[1]
\REQUIRE $R$, The set of attributes distributed among the parties $P_{ij}$ with $i=1,..,\vb{v}$,  $j=1,..,\vb{h}$. 
\REQUIRE $C$, The class attribute with $d$ class values, $c_1,...,c_d$. 
\REQUIRE $S$, the grid distributed data set over parties $P_{ij}$ with $i=1,..,\vb{v}$,  $j=1,..,\vb{h}$ and parties $P_{\vb{v},j}$ holding the class attribute . 
\IF{(Emptiness test)The parties test if $R$ is empty} 
\STATE Secure sum protocol and Yao circuit are used to test whether $R$ is empty.
\STATE In case the attribute set is empty, Secure size of set intersection protocol, secure sum protocol and Yao circuit are used to calculate which class value $c_{i}$ is most frequent. a leaf node with class value $c_{i}$ is returned.

\ELSIF{All parties test whether all tuples have the same class value $c_{i}$} 
\STATE Secure size of set intersection protocol, secure sum protocol and Yao circuit are used to calculate this.
\STATE In case the test is TRUE, a leaf with class value $c_{i}$ is returned.  
\ELSE 
\STATE Determine attribute $A$, classifying most accurately tuples in $S$: use secure size of intersection, secure sum and $xln(x)$ protocols. 
\STATE Partition $S$ in $m$ parts $S(a_{1}), ..., S(a_{m})$ such that $a_{1} ... a_{m}$ are the different values of $A$.
\STATE Return a tree with root $A$ and $m$ branches $a_{1} ... a_{m}$ such that branch $i$ contains $ID3(R - \left\{A\right\},C,S(a_{i}))$. 
\ENDIF
\end{algorithmic}
\end{algorithm}

\section{Complexity analysis of privacy preserving ID3 over grid-partitioned data}
\label{sec:discussion}

In this section we analyse the complexity of the two computation strategies proposed in the previous section:
 first merging horizontally and developing vertically or 
first merging vertically and next developing horizontally.
 
The different quantities  $h$, $v$,$k$, $|T|$, $|R|$, $\cold$, $\iold$, $t$ and $n$ that play a role in this analysis are explained in the next table.
 
 \medskip
\begin{center}
\begin{tabular}{|c | l |}
\hline
{\bf Notation}& {\bf Meaning}\\
\hline\hline
$h$& the number of horizontal groups \\
$v$& the number of  vertical groups\\
$k$& the number of parties \vb{(=$h \times v$)}\\
$|T|$& the number of tuples in the data set\\
$|R|$& the number of attributes\\
$\cold$& the number of values for the class attribute $C$\\
$\iold$& the maximal number of values for an attribute\\
$t$& the maximal length of encryption keys\\
$n$& the maximal length of Taylor series\\

\hline
\end{tabular}
\end{center}
\medskip

The predominant task in the ID3 algorithm is to determine the attribute with the highest Information Gain and we will base our analysis mainly on this component. 

\subsection{The complexity of the components from SMPC}

In discussing the complexity of the building blocks described in Section~\ref{subsec:smpc}
usually two components are considered: the \emph{computational complexity} and the \emph{communication complexity}. The former considers the cost of computations in the classical sense, the latter considers the cost of passing messages between, e.g., between different parties.

 \subsubsection{The complexity of the secure sum protocol}

For the secure sum protocol with $k$ parties, the computation and communication costs are both
$O(k\log{(|T|)})$. Each of the parties never outputs values larger than $|T|$ and the messages passed are never larger than $|T|$. Assuming binary encoding of numbers, 
this gives the above result. 

 \subsubsection{The complexity of the secure union protocol and secure size of intersection protocol}

For the secure union protocol and the secure size of intersection protocol with $k$ parties, the computation cost is 
$O(k^2|T|t^3)$ and the communication cost is $O(k^2|T|t)$. 
Indeed, the parties send sets of at most size $|T|$, they make use of encryption keys of length 
 $t$, hence the factor $t^3$ in the computation cost, and 
 every party has to encrypt $k^2$ sets. 
The value $t$ in the communication cost points at the size of the sets that are transmitted. In total $k^2$ messages are sent of size 
$|T|t$.  

 \subsubsection{The complexity of the secure $x\ln{x}$ protocol and Yao circuits}
 
 The secure $x\ln{x}$ protocol for two parties has a computational cost of $O(\log{(|T|)})$ and a communication cost of $O(n\log{(|T|)}t) $.
 It takes input values of at most $|T|$.
 The protocol also depends on a value $n$ that determines how far a Taylor series is 
 developed. The protocol contains a  Yao circuit, created by one of the parties who also gives his input to the circuit and passes it to the other party. This explains the communication cost, in which $n$ obviously plays a role since it determines the size of the circuit.
 The second party receives the circuit and feeds his input to the circuit.
 Hereto one \vb{oblivious} transfer is performed per bit. This step explains the computational cost.
 
 \subsection{The complexity of first horizontal merging}

To determine the attribute that best classifies the data, for each attribute unions have to be determined over $h$ parties.
The exact number of unions may depend on the attribute under consideration.
If it is an attribute that belongs to the party that possesses the class attribute, 
it are  $v + \cold + \iold -1$ unions. If an other party belongs the class attribute, it are 
$v + \cold + \iold - 2$ unions. Also the number of unions to be transmitted depends on the attribute.
For attributes in the possession of the owner of the class attribute, there are $v -1$ unions to be transmitted, for other attributes $v + \cold - 2$. \\
So, we conclude: 
\medskip
\par\noindent 
\indent{\bf computation cost} = $O(|R|(v + \cold + \iold)(h ^2|T|t^3))$ \\
\ \\
{\bf communication cost} = $O(|R|(v + \cold)(h^2|T|t))$.

We end this section with a remark on how this protocol could be made more efficient.
Remark that the strength of this protocol resides in the fact that adjacent parties may use the same encryption key. For this reason it is not necessary to use the 
secure size of set intersection protocol to calculate intersections.
The first phase of this protocol can be skipped because the same keys are used 
when computing unions.
This is possible here because the data is both horizontally and vertically distributed.
When the data is only vertically distributed, it would also be possible to let the parties
agree on some encryption keys and to simplify the protocol in this way.

\subsection{The complexity of first vertical merging}
Per attribute  $1 + \cold + \iold + \cold\iold$ values have to be computed, namely:
\begin{itemize}
\item
The number of transactions reaching the current node: 1; 
\item
The number of transactions reaching the current node per class value: $\cold$. 
\item 
The number of transactions reaching the current node per \vb{attribute} value: $\iold$. 
\item
The number of transactions reaching the current node per class value and per attribute value: $\cold\iold$.
\end{itemize}

All these values can be computed via a secure size of set intersection protocol that is each time executed by $v$ parties.
Since we have to count this for each horizontal group, this gives in total $h(1 +\cold +\iold +\cold\iold)$ 
calls to the  secure size of set intersection protocol. 
With these values the computation continues. In case of two horizontal groups this is with the $x\ln{x}$ protocol; in the case of more horizontal groups this is with the secure sum protocol, followed by the $x\ln{x}$ protocol.
\noindent So, we conclude: \\
\medskip
\begin{tabular}{rl}
{\bf computation cost} =&$O(|R|(h(1 +\cold +\iold +\cold\iold))(v^2|T|t^3)+$\\ 
&$|R|(1 +\cold + \iold + \cold\iold)(\log{(|T|)}) +$\\ 
&$ |R|\log{(|T|)}) \ \ [+O(h\log{(|T|)})]$ 
\end{tabular}\\

and

\begin{tabular}{rl}
{\bf communication cost} =&$O(|R|(h(1 + \cold + \iold + \cold\iold)).(v^2|T|t)+$\\
&$|R|(1 + \cold + \iold + \cold\iold)n(\log{(|T|)}t)+$\\
&$|R|\log{(|T|)}t)  \ \ [+O(h\log{(|T|)})] $.
\end{tabular}

 \subsection{Conclusion on the  complexity analysis}
 
 We start by remarking that it looks more logical to merge the data first horizontally and then to further develop it vertically. De emptiness test can be implemented more efficiently in the former case.
 The secure $x\ln{x}$ protocol gives an approximated result, but the difference from the real result  is small. This protocol also makes heavy use of circuit computations. In practice it is preferable to  avoid this.
 
For what concerns complexity, the above obtained expressions also show that 
first horizontally merging is advantageous.
 And as remarked before it can be improved by an  optimal use of encryption. Indeed, by giving different parties the same encryption key it is not necessary to perform the 
 secure size of set intersection protocols after the 
  secure union protocols have been executed.
  
\section{Conclusions} \label{sec:conclusion}
In this paper we first discussed the significance of extending the current state of the art in privacy preserving datamining to grid partitioned data, i.e. data which is as well horizontally as vertically partitioned. Our motivating example shows that this situation is of great interest to real world situations and applications. Then we continued by formally defining horizontally, vertically and grid partitioned data. To our knowledge we are the first to formalize the concept of grid partitioned data.

We continued by introducing three new privacy preserving data mining algorithms. We started by extending the result of Lindell and Pinkas \cite{Lind01}, i.e. preserving privacy for decision tree learning with two parties when data is horizontally distributed, to more than two parties. However, the main contribution of this paper are the two algorithms to securely induce a distributed decision tree when data is grid partitioned. More precisely, we considered two possible solutions: one in which data is first merged horizontally and then further developed vertically and vice versa. The complexity analysis of both algorithms shows that it is more efficient to first merge data horizontally and further develop it vertically than the other way around.

\bibliographystyle{elsart-num-sort}

\end{document}